# Experimental realization of a continuous version of the Grover algorithm

Vladimir L. Ermakov[1] and B. M. Fung[2]

*Department of Chemistry and Biochemistry, University of Oklahoma,*

*Norman, Oklahoma 73019-0370, USA.*

[1]On leave from *Kazan Physical-Technical Institute, Kazan 420029, Russia;* e-mail addresses: ermakov@ou.edu (current), ermakov@smtp.ru (permanent).

[2]Corresponding author; e-mail address: bmfung@ou.edu.

## Abstract

A continuous, analog version of the Grover algorithm is realized using NMR. The system studied is $^{23}$Na in a liquid crystal medium. The presence of quadrupolar coupling makes the spin $I=3/2$ nucleus a 2-qubit system. Applying a specially designed pulse sequence, the time evolution of the spin density operator is described in an interaction representation which has no external time-dependent radio-frequency fields. This approach is used to implement one instance of the continuous Grover search for the transform of a uniform state to a target state, and the implementation provides a clear physical interpretation of the algorithm. The experimental results are in good agreement with the theory.

# I. INTRODUCTION

The idea of quantum computer was first proposed by Richard Feynman in 1982 [1]. He noted that simulating quantum dynamics using classical computers is intrinsically difficult, because the memory space and processing time required to perform the simulation grows exponentially with the size of the quantum system to be simulated. He suggested that this difficulty might be overcome by performing quantum simulations on quantum simulators, which are devices whose memory space and processing time grow only as a small polynomial in the size of the quantum system to be simulated. No one was sure how to use the quantum effects to speed up the computation until 1994, when Peter Shor discovered a polynomial time quantum algorithm for factoring integers [2].

Feynman's quantum computer [3] essentially operates as an analog computer: it is a quantum system whose dynamics can be programmed to mimic the dynamics of the quantum system of interest. However, most quantum algorithms have been realized using algorithmic approach, when a physical system works like a digital quantum computer. The Grover algorithm [4] is an important digital quantum algorithm, and is the basis of quantum search. Its first analog analogue was described by Farhi and Gutmann [5]. Later Fenner found a time-independent Hamiltonian for a system of quantum bits that results in time evolution matching Grover's iteration exactly [6].

Several different techniques have been used to construct prototype quantum computers [7]. Among these, NMR is the best developed experimental approach [7-11]. It has been used to carry out a number of quantum simulations and quantum algorithms, including the factorization of a number (15=3×5) using the Shor algorithm [11]. In general, to carry out quantum information processing by NMR, a pseudopure (effective pure) spin state [12,13] is first prepared, followed by the application of a series of pulses with delays related to the coupling constants to perform unitary operations. However, it has been pointed out [5] that the Grover quantum search can be achieved by a continuous evolution of an initial state under the influence of a time-independent Hamiltonian. We have realized such a method experimentally, and the results are reported here. The NMR system used is a spin 3/2 nucleus showing well-defined quadrupole splitting. Nuclei with I=3/2 in ordered environments can form 2-qubit systems [14-16], and nuclei with I=7/2 can form 3-qubit systems [17-19].

## II. THE FARHI-GUTMANN-FENNER APPROACH

To be specific, let us consider a 2-qubit system with four marked pseudopure states and a uniform state |s>, which is a superposition of the four marked states:

$$|w_0> = \begin{pmatrix} 1 \\ 0 \\ 0 \\ 0 \end{pmatrix}, \quad |w_1> = \begin{pmatrix} 0 \\ 1 \\ 0 \\ 0 \end{pmatrix}, \quad |w_2> = \begin{pmatrix} 0 \\ 0 \\ 1 \\ 0 \end{pmatrix}, \quad |w_3> = \begin{pmatrix} 0 \\ 0 \\ 0 \\ 1 \end{pmatrix}, \quad |s> = \frac{1}{2}\begin{pmatrix} 1 \\ 1 \\ 1 \\ 1 \end{pmatrix}. \quad (1)$$

In the following, the binary notations $|w_0>=|00>$, $|w_1>=|01>$, $|w_2>=|10>$, and $|w_3>=|11>$ are used for the pseudopure states.

Farhi and Gutmann proposed [5] that a time-independent Hamiltonian can perform a quantum search in a database with N items. Their Hamiltonian is the sum of two simple Hamiltonians, an "oracle" |w><w| and a "driver" Hamiltonian |s><s|. The oracle Hamiltonian has

a large (N−1)-fold degenerate ground state and a single "marked" excited state |w>, and the driver has a large (N−1)-fold degenerate ground state and a single excited state |s>. The quantum system begins in the state |s> and is reflected into a "marked" state |w> after a time $O(\sqrt{N})$. Fenner showed [6] that, although the evolution under the Farhi-Gutmann Hamiltonian reaches the same final state as the Grover algorithm does, the time evolution of the system according to the Hamiltonian strays far from the intermediate steps in Grover's algorithm; nevertheless, the commutator of an "oracle" and a "driver" gives the adequate Hamiltonian.

The Fenner Hamiltonian for our system is (details that can be found in [6] are omitted):

$$H_f = 2\,\Omega_f\, i\, (\,|w_2><s| - |s><w_2|\,) = \Omega_f\, i \begin{pmatrix} 0 & 0 & -1 & 0 \\ 0 & 0 & -1 & 0 \\ 1 & 1 & 0 & 1 \\ 0 & 0 & -1 & 0 \end{pmatrix}, \qquad (2)$$

where $\Omega_f$ is an adjustable amplitude. It was shown that, if the time interval $t_f$ is such that

$$2\,\Omega_f\, t_f = 2[\pi - 2\cos^{-1}(0.5)]/\sqrt{3} \cong 1.2092, \qquad (3)$$

the initial uniform state is transformed to the marked state, i.e., $\exp(-i\,H_f\,t_f\,)\,|s> \Rightarrow |w_2>$.

### III. IMPLEMENTATION

The system studied is $^{23}$Na, which has spin I=3/2, in a liquid crystalline medium. For this system in the presence of a large magnetic field, the Hamiltonian can be written as [20] (throughout the paper we use $\hbar = 1$):

$$H_0 = -\omega_Z I_z + \omega_Q [I_z^2 - I(I+1)/3]/4. \tag{4}$$

When $\omega_Z \gg \omega_Q = e^2qQ \cdot P_2(\cos\theta)$, where the second order Legendre polynomial $P_2(\cos\theta)$ is the order parameter, the quadrupole interaction acts as a perturbation to the Zeeman interaction, so that

$$H_0 = \begin{pmatrix} -\frac{3}{2}\omega_Z + \frac{1}{4}\omega_Q & 0 & 0 & 0 \\ 0 & -\frac{1}{2}\omega_Z - \frac{1}{4}\omega_Q & 0 & 0 \\ 0 & 0 & +\frac{1}{2}\omega_Z - \frac{1}{4}\omega_Q & 0 \\ 0 & 0 & 0 & +\frac{3}{2}\omega_Z + \frac{1}{4}\omega_Q \end{pmatrix}$$

$$= \begin{pmatrix} \varepsilon_0 & 0 & 0 & 0 \\ 0 & \varepsilon_1 & 0 & 0 \\ 0 & 0 & \varepsilon_2 & 0 \\ 0 & 0 & 0 & \varepsilon_3 \end{pmatrix}. \tag{5}$$

The general form of the radio frequency (RF) Hamiltonian can be written as

$$H_1(t) = \omega_1(t) I_Y, \tag{6}$$

where $\omega_1(t)$ is a time dependent function and $I_Y$ is the observable operator for this system expressed by the matrix

$$I_Y = \begin{pmatrix} 0 & -i\frac{\sqrt{3}}{2} & 0 & 0 \\ i\frac{\sqrt{3}}{2} & 0 & -i & 0 \\ 0 & i & 0 & -i\frac{\sqrt{3}}{2} \\ 0 & 0 & i\frac{\sqrt{3}}{2} & 0 \end{pmatrix}. \tag{7}$$

Our goal is to construct the Fenner Hamiltonian (2) using an adequate choice of the function $\omega_{rf}(t)$. For the I=3/2 spin system, the four energy levels are not equally spaced, so that six different frequencies can be applied to excite three single quantum transitions, two double quantum transitions, and one triple quantum transition. As a consequence, the RF operator does not need to preserve its original form (7) and can be shaped to have, in principle, any matrix elements (in our system the triple quantum transition frequency is the same as the frequency of the central single quantum transition, so that two of the matrix elements cannot be easily created).

To analyze one of several possibilities, let us consider a three-frequency RF Hamiltonian:

$$H_1 = 2\left[\Omega_{02} \cos(\omega_{02} t + \varphi_{02}) + \Omega_{12} \cos(\omega_{12} t + \varphi_{12}) + \Omega_{23} \cos(\omega_{23} t + \varphi_{23})\right] I_Y. \tag{8}$$

where $\Omega_{ij}$, $\omega_{ij}$ and $\varphi_{ij}$ are amplitudes, frequencies, and phases of the three harmonics, respectively. Here the frequencies are chosen to fulfill the resonant conditions $\omega_{12} = \varepsilon_1 - \varepsilon_2$, $\omega_{23} = \varepsilon_2 - \varepsilon_3$ (single quantum transitions) and $\omega_{02} = (\varepsilon_0 - \varepsilon_2)/2$, (double quantum transition). For further consideration, the RF Hamiltonian (8) is split into a secular time-independent part and a non-secular oscillating part (here and in the following we do not use any special notation for operators in the interaction representation defined by $H_0$):

$$H_1(t) = H_1^{sec} + H_1^{nonsec}(t). \tag{9}$$

where the time-independent terms are

$$H_1^{sec} = <1|H_1^{sec}|2> + <2|H_1^{sec}|1> + <2|H_1^{sec}|3> + <3|H_1^{sec}|2> = \\ = \Omega_{12}(<1|I_Y|2> + <2|I_Y|1>) + \Omega_{23}(<2|I_Y|3> + <3|I_Y|2>). \tag{10}$$

Because the matrix element <0|$I_Y$|2> (as well as <2|$I_Y$|0>) vanishes, and the time-dependent RF perturbation cannot cause any transitions from |0> to |2> in the first order, the amplitude of the transition is determined by the second order matrix element. The general formula for the intensity of NMR multiple-quantum transitions can be obtained form perturbation theory [21], and the expression for the double quantum transition in spin 3/2 systems is quite simple [22]. Experimentally, we used a calibration procedure to determine the value of the RF amplitude $\Omega_{02}$ which gives the right value $\Omega_{02}^{eff}$. When the condition is fulfilled and the pulse frequencies are set to the correct values of $\omega_{12}$, $\omega_{23}$, and $\omega_{02}$, respectively, the RF Hamiltonian takes the time-independent form:

$$H_1 = \begin{pmatrix} 0 & 0 & -i\Omega_{02}^{eff} & 0 \\ 0 & 0 & -i\Omega_{12} & 0 \\ i\Omega_{02}^{eff} & i\Omega_{12} & 0 & -i\frac{\sqrt{3}}{2}\Omega_{23} \\ 0 & 0 & i\frac{\sqrt{3}}{2}\Omega_{23} & 0 \end{pmatrix}. \tag{11}$$

To finally shape the Hamiltonian (11) into the Fenner Hamiltonian form (2), it is necessary to adjust the laboratory frame RF amplitudes $\Omega_{02}$, $\Omega_{12}$ and $\Omega_{23}$ in such a way that the magnitude of all non-zero matrix elements in (11) are equal to the Fenner constant $\Omega_f$:

$$\Omega_{02}^{eff} = \Omega_{12} = \frac{\sqrt{3}}{2}\Omega_{23} \equiv \Omega_f. \qquad (12)$$

This adjustment was accomplished by a calibration procedure. Three independent experiments were performed. In each of them the duration of a pulse was fixed to $t_{\pi/2} = 2000$ μs and only one frequency was excited. When irradiating only one transition the system can be considered as effective spin 1/2 [23, 24] with the usual Pauli matrix transformation rule:

$$\exp(-i\,\omega t\,\sigma_y)\,\sigma_z\,\exp(+i\,\omega t\,\sigma_y) = \cos(2\omega t)\,\sigma_z + \sin(2\omega t)\,\sigma_x, \qquad (13)$$

One starts with the equilibrium state $\sigma_z$. Theoretically, either the duration or the strength of the RF pulses must be proportional to $\sqrt{N}$. In practice, the pulse amplitude can only be calibrated experimentally because it is determined by the real RF power, but the RF amplifier is not truly linear and the tuning of the probe is sample-dependent. Therefore, the RF power for any specific frequency is adjusted to produce the state $\sigma_x$ which gives the maximum signal, and it gives the relation $2\Omega_f\,t_{\pi/2} = \pi/2$. In this way, it was found that the values $\Omega_{12} = 62.5$ Hz, $\Omega_{23} = 72.2$ Hz, and $\Omega_{02} = 458.5$ Hz gave the same $\Omega_f$.

By modulating the phase and amplitude of the RF carrier, the three different frequencies with the amplitudes specified above can be applied simultaneously to fulfill the condition set in Eq. (8). Then, equal matrix elements in the representation given in Eq. (11) can be created. The Fenner time $t_f$, which is necessary to achieve the full coherence transfer between the uniform state and a target state, can be calculated as follows. In addition to the condition $\Omega_f t_{\pi/2} = \pi/4$ used during the calibration, one can see from Eq. (3) that the Fenner time satisfies $2\Omega_f t_f \approx 1.2092$. Knowing the length of the $\pi/2$ pulse, $t_{\pi/2} = 2000$ μs, one gets the value $t_f \approx t_{\pi/2} 1.2092/(\pi/2) \approx 0.7698 \cdot t_{\pi/2} = 1540$ μs. This value was used for the Grover pulses.

The experimental procedure is the following. First, one of the pseudopure states, the marked state $|w_1\rangle$, is produced by using a double-quantum 90° pulse followed by a selective 180° single quantum pulse with phase cycling [15]. Second, an inverse Grover iteration is applied to transform $|w_1\rangle$ to the uniform state $|s\rangle$. Third, a direct Grover iteration is applied to transform $|s\rangle$ into the marked state $|w_2\rangle$. A continuous Hamiltonian of the form (11) is used for both the direct Grover iteration, Eq. (2), in the third step, and the inverse iteration in the second step.

## IV. EXPERIMENTAL RESULTS

The sample used was a lyotropic liquid crystal composed of 37% decylsulfate, 7% decanol, and 56% water. The NMR experiments were carried out using a Varian UNITY/INOVA 400 spectrometer at 22 °C, with $^{23}$Na resonance frequency at 105.79 MHz. The $^{23}$Na quadrupole splitting (the full spectrum width) $\omega_Q/2\pi$ was 10840 Hz. $T_1$=16 ms; $T_2$ was 16 ms for the central peak and 4.5 ms for the outside peaks.

Our experimental procedure can be represented schematically by consecutive transformations: [system in thermal equilibrium] $\Rightarrow |00\rangle \Rightarrow |01\rangle \Rightarrow |s\rangle \Rightarrow |10\rangle$, where the last step ($|s\rangle \Rightarrow |10\rangle$) corresponds to one of four possible Grover iterations. The schematic diagram for the pulse sequence is shown in Fig. 1 (row I). The diagram shows the real pulse shapes – the Gaussian shapes for the selective pulses and the rectangular shapes for the three Grover pulse components and monitoring pulse – but all amplitudes are shown approximately equal to fit the picture (the real amplitude of the Grover pulse double quantum component is about 7 times more that that of single quantum components, as discussed in the previous section).

1. The first step is the preparation of the ground pseudopure state |00> from the equilibrium state using a double quantum 90º pulse (DQ-90º) with appropriate phase cycling [14], Fig.1 (a, I). The pulse has a Gaussian shape, and the length was 2.00 ms.

2. The second step is the transformation of this pseudopure ground state to another pseudopure state $|w_1>=|01>$ using a selective 180º single quantum pulse (SQ-180º), Fig. 1 (b, I). The pulse has a Gaussian shape, and the length was 1.50 ms.

3. The next step is the transformation of $|w_1>=|01>$ state to the uniform state |s> using the inverse Grover iteration, Fig. 1 (c, I). The pulse length was 1.54 ms.

4. The last step is the transformation of the uniform state |s> to the marked state $|w_2>=|10>$ using direct Grover iteration, Fig. 1 (d, I). The pulse length was also 1.54 ms.

Fig. 1 also shows the vector notation of the spin state after each step (row II), the scheme of excess population on the four energy levels (row III), and the expected spectrum (row IV) after the application of a non-selective $\pi/20$ pulse to monitor the populations on all energy levels of each state.

The corresponding experimental spectra are shown in Fig. 2. For the system in equilibrium, Fig. 2(a), the integrated intensity ratios of the three peaks are 3:4:3; the outer peaks are considerably broadened so that their peak heights are lower. The final spectrum, Fig. 2 (e), has a reduction on the signal intensity due to spin-lattice relaxation. Taking these into account, the experimental results show excellent agreement with the theoretical spectra depicted in Fig. 1 (row IV).

We have presented the results of only one of four possible Grover iterations, $|s\rangle \Rightarrow |10\rangle$. Another Grover iteration, namely $|s\rangle \Rightarrow |01\rangle$, is the direct version of the inverse Grover iteration ($|01\rangle \Rightarrow |s\rangle$) used in the second step. The result (not shown here) is similar to Fig. 2; the only difference is that the last spectrum (e), is replaced by spectrum (c) with slightly reduced amplitude (due to spin-lattice relaxation). To construct the two other possible Grover iterations, $|s\rangle \Rightarrow |00\rangle$ and $|s\rangle \Rightarrow |11\rangle$, the triple quantum transition must be excited. As mentioned in Sec. III, in our case the frequency for triple quantum transition, $\omega_{03} = (\varepsilon_0 - \varepsilon_3)/3$, coincides with that of the central single-quantum transition, $\omega_{12} = (\varepsilon_1 - \varepsilon_2)$ [22, 24]. It is necessary to use complicated schemes to excite only the former without affecting the latter. Because of this difficulty, we have not carried out such experiments.

## V. SUMMARY

We have shown that the Grover quantum algorithm can be realized in an adequate interaction representation by a time-independent Hamiltonian with all parameters well controlled by the experimentalist. In summary, starting with one of the pseudopure states, Fig. 1(b) and Fig. 2(c), the superposition of all quantum states was prepared by using an inverse Grover operation, Fig. 1(c) and Fig. 2(d). This superposition has proper non-zero off-diagonal elements in the density matrix, and is entirely different from a state in which equal populations are prepared by complete RF saturation. By applying a continuous Grover operation on this uniform superposition of quantum states, another pseudopure state is generated, Fig. 1(d) and Fig. 2(e), achieving the purpose of Grover search.

ACKNOWLEDGEMENT

This work was supported by the National Science Foundation under grant numbers DMR-9809555 and DMD-0090218.

## Figure captions

**Fig. 1**. (**I**) Schematic pulse sequence: (**a**) double quantum 90° pulse; (**b**) selective 180° pulse; (**c**) three-frequency inverse Grover pulse; (**d**) three-frequency direct Grover pulse; (**e**) monitoring $\pi/20$ pulse. (**II**) The corresponding spin states in vector notation. (**III**) The spin states in a notation showing excess populations on the four energy levels. It should be noted that state (c) is not diagonal and has off diagonal matrix elements. (**IV**) Expected spectra after applying a monitoring pulse after each step.

**Fig. 2**. Experimental spectra recorded after applying the following pulse sequences to the system in the thermal equilibrium. (a): $(\pi/20)$; (b): (DQ-90°)–$(\pi/20)$; (c): (DQ-90°)–(SQ-180°)–$(\pi/20)$; (d): (DQ-90°)–(SQ-180°)–(GROVER$^{-1}$)–$(\pi/20)$; (e): (DQ-90°)–(SQ-180°)–(GROVER$^{-1}$)–(GROVER)–$(\pi/20)$. They correspond to the schematic expected spectra in Fig. 1 (row IV).

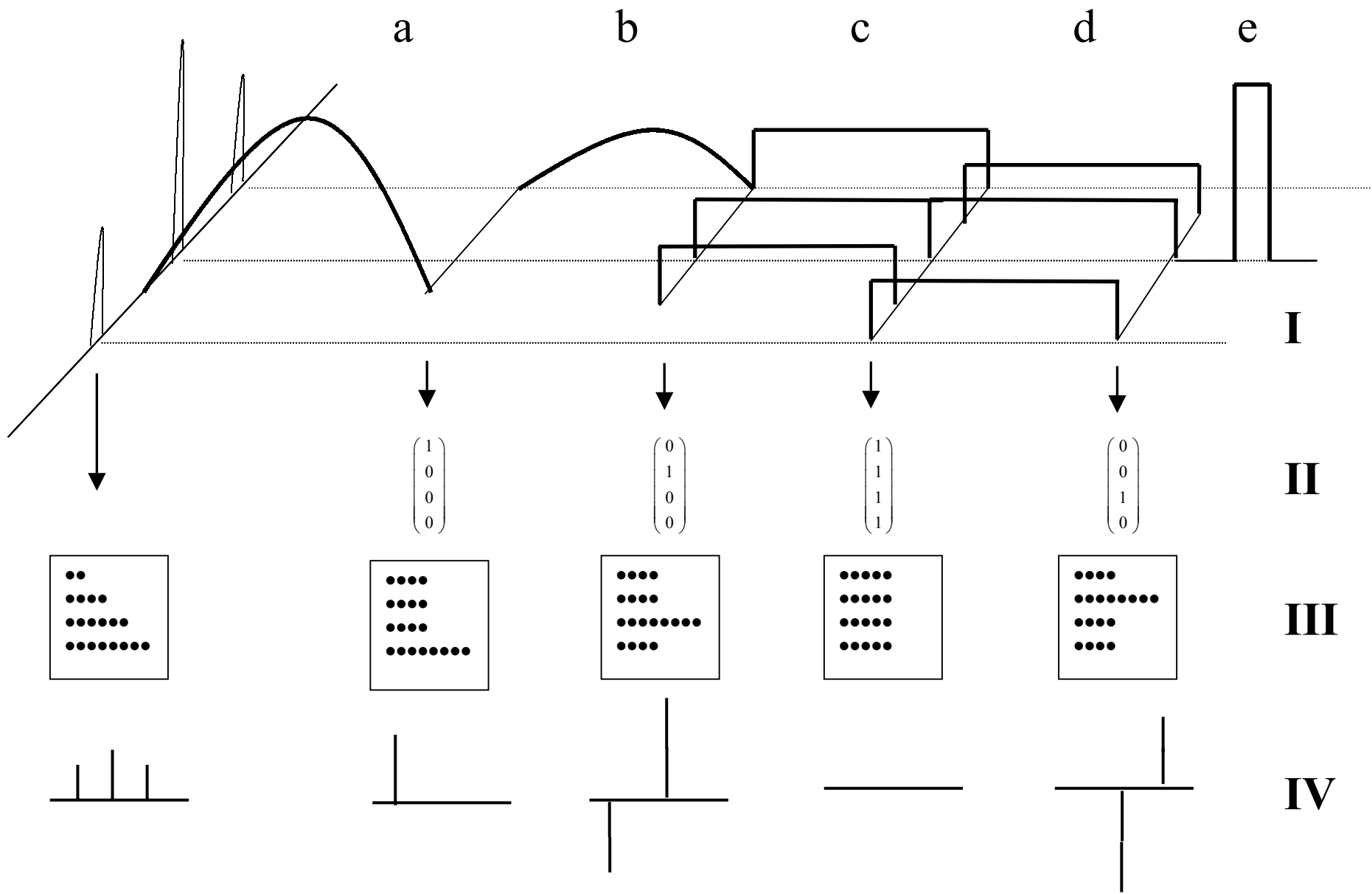

Fig. 1, Ermakov, Phys. Rev. A

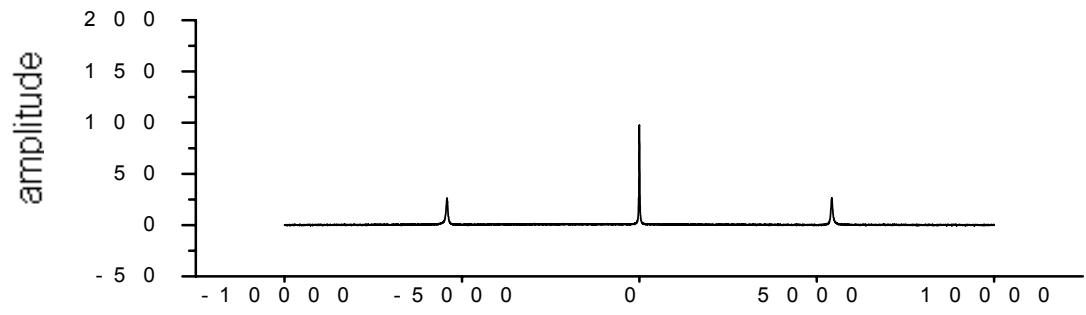
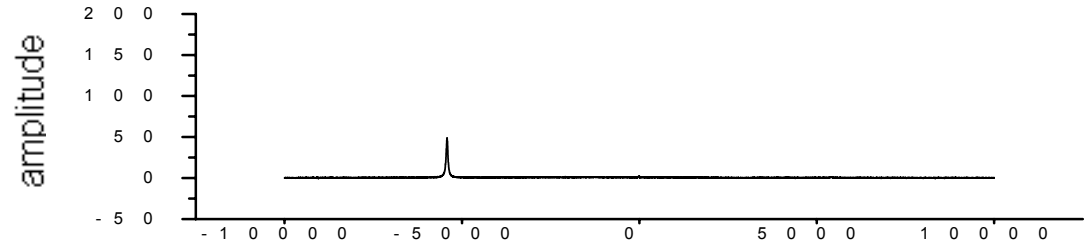
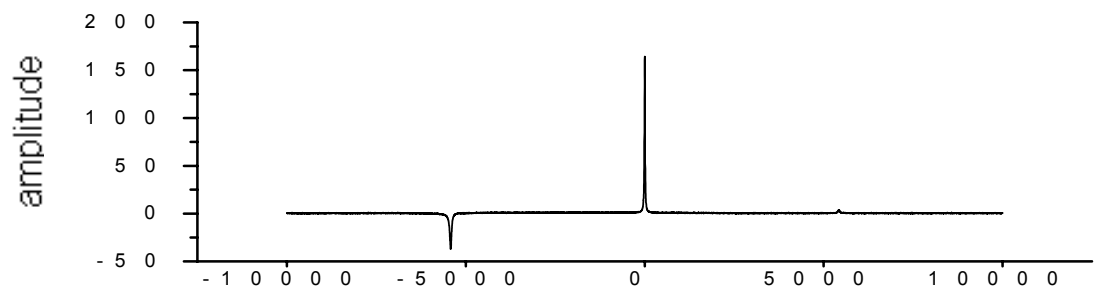
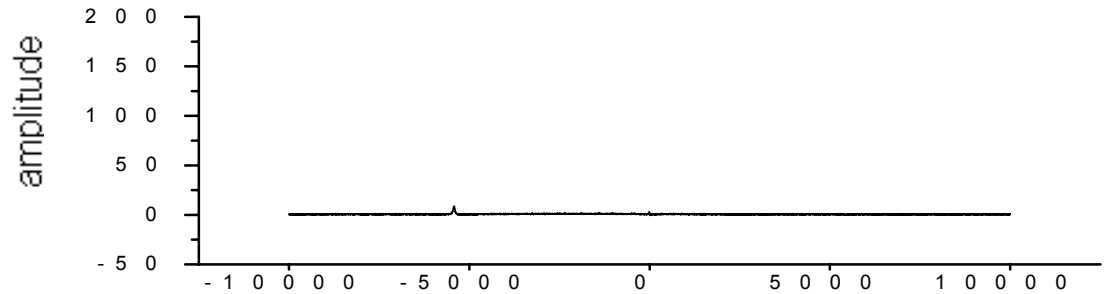
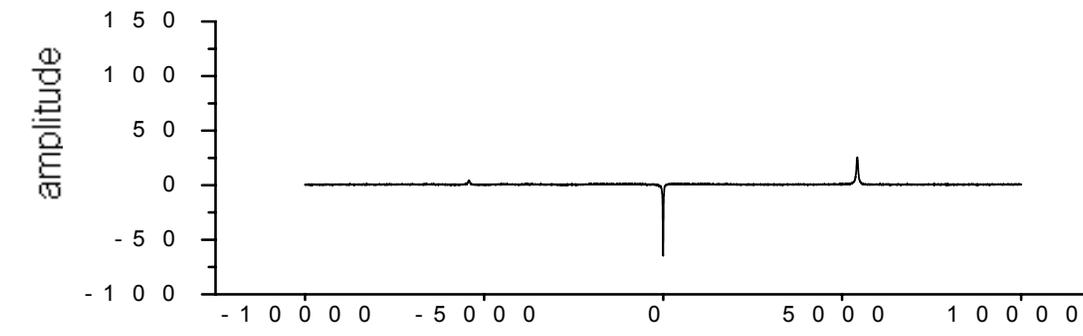

Fig. 2 Ermakov, Phys. Rev. A